\documentclass[twocolumn,prl,preprintnumbers,superscriptaddress,amsmath,amssymb,english]{revtex4}
\usepackage{amsmath,amssymb,amsfonts,mathrsfs}
\usepackage{amsthm}
\usepackage{CJK}
\usepackage{bm}
\usepackage{babel}
\usepackage{graphicx}
\allowdisplaybreaks
\usepackage{braket}
\usepackage[breaklinks]{hyperref}
\usepackage{color}
\hypersetup{colorlinks=true, linkcolor=blue, citecolor=blue, filecolor=blue, urlcolor=blue}

\begin{document}

\title{A Photoinduced Floquet Mixed-Weyl Semimetallic Phase in a Carbon Allotrope}
\author{Tingwei Deng}
\affiliation{Institute for Structure and Function $\&$ Department of Physics, Chongqing University, Chongqing 400044, P. R. China}

\author{Baobing Zheng}
\affiliation{College of Physics and Optoelectronic Technology, Nonlinear Research Institute, Baoji University of Arts and Sciences, Baoji 721016, P. R. China}
\affiliation{Institute for Structure and Function $\&$ Department of Physics, Chongqing University, Chongqing 400044, P. R. China}

\author{Fangyang Zhan}
\affiliation{Institute for Structure and Function $\&$ Department of Physics, Chongqing University, Chongqing 400044, P. R. China}

\author{Jing Fan}
\affiliation{Center for Computational Science and Engineering, Southern University of Science and Technology, Shenzhen 518055, P. R. China}

\author{Xiaozhi Wu}
\affiliation{Institute for Structure and Function $\&$ Department of Physics, Chongqing University, Chongqing 400044, P. R. China}

\author{Rui Wang}
\email{rcwang@cqu.edu.cn}
\affiliation{Institute for Structure and Function $\&$ Department of Physics, Chongqing University, Chongqing 400044, P. R. China}
\affiliation{Center for Quantum materials and devices, Chongqing University, Chongqing 400044, P. R. China}

\begin{abstract}
The interplay between light and matter attracts tremendous interest for exploring novel topological quantum states and their phase transitions. Here we show by first-principles calculations and the Floquet theorem that a carbon allotrope bct-C$_{16}$, a typical nodal-line semimetal, exhibits exotic photoinduced Floquet mixed-Weyl semimetallic features. Under the irradiation of a linearly polarized light, bct-C$_{16}$ undergos a topological phase transition from a Driac nodal-line semimetal to a Weyl semimetal with two pairs of tunable Weyl points. With increasing the light intensity, left-handed Weyl points evolve from type-I into type-II while right-handed ones are always preserved to be type-I, giving rise to photo-dressed unconventional Weyl pairs composed of distinct types of Weyl points. Importantly, a special Weyl pair formed by type-I and type-III Weyl points is present at the boundary between type-I and type-II states. The Floquet Fermi arcs connecting the projections of two different types of Weyl points are clearly visible, further revealing their unique topological features. Our work not only realizes promising unconventional Weyl pairs but also paves a reliable avenue for investigating light-induced topological phase transitions.
\end{abstract}

\pacs{73.20.At, 71.55.Ak, 74.43.-f}

\keywords{ }

\maketitle

Topological semimetals  possess topologically protected fermionic quasiparticles, extending the topological classification of condensed mater beyond insulators due to their nontrivial electronic wave functions \cite{Kane-RevModPhys.82.3045, ZSC-RevModPhys.83.1057,RevModPhys.90.015001}. For these materials, the valence and conduction bands cross near the Fermi level in the momentum space, forming the point-like or line-like Fermi surfaces \cite{RevModPhys.90.015001}. Accordingly, various topological fermions, such as Dirac fermions \cite{WangPhysRevB.85.195320, ScienceLiu864}, Weyl fermions \cite{Wan2011,Xu2011,PhysRevX.5.011029,NatureSoluyanov2015}, nodal-line fermions \cite{PhysRevB.84.235126,PhysRevLett.115.036806,PhysRevLett.115.036807, PhysRevB.97.241111}, triple fermions \cite{ZhuPhysRevX.6.031003,lvNature2017}, as well as beyond \cite{PhysRevLett.116.186402,Hourglass,Bradlynaaf5037}, have been proposed. Most of them have been verified in experiments \cite{ScienceLiu864, lvNature2017, Xu613,Lv2015,Morali1286, Belopolski1278,Liu1282}. Among these nontrivial fermionic quasiparticles, Weyl fermions in Weyl semimetals (WSMs) are of particular importance. The nodal point [i.e., Weyl point (WP)] in WSMs is characterized by specific chirality (right- or left-handed), acting as a topological monopole in the field of Berry curvature. According to the manifold of the Fermi surface, two types of Weyl fermions have been identified \cite{PhysRevLett.114.016806, PhysRevB.91.115135,NatureSoluyanov2015}. The first type (i.e., type-I Weyl fermions) corresponds to a standard Weyl cone associating with the Lorentz invariance, which is characterized by a closed isoenergetic contour eventually evolving into a point-like Fermi surface. The second type (i.e., type-II Weyl fermions) corresponds to a tilted Weyl cone, which is characterized by an open Fermi surface with two crossing isoenergetic contours. For the type-II Weyl fermions, the Lorentz invariance is broken. Furthermore, a particularly interesting situation occurs at the boundary between type-I and type-II Weyl cones. This critical transition point is accompanied by a flat band along one direction, termed as a type-III WP \cite{PhysRevLett.114.016806, PhysRevLett.120.237403, PhysRevB.98.121110,PhysRevX.9.031010}. For the type-III Weyl fermions, the Fermi surface is a single line, inducing highly anisotropic effective masses \cite{PhysRevX.9.031010}. More importantly, the type-III WPs offer the possibility to the study of the event horizon of a black hole in crystalline solids \cite{PhysRevB.101.035130}.

A significant hallmark of WSMs is the nontrivial Fermi arc \cite{Wan2011}, which connects the projections of two WPs with opposite chirality on a surface. These two WPs form a Weyl pair. Usually, a conventional Weyl pair contains two same types of WPs. The presence of paralleled electric and magnetic fields can switch the number of paired Weyl fermions with opposite chirality, inducing that the classical conservation of topological charge is broken in a Weyl system. This effect is known as the chiral anomaly \cite{RevModPhys.90.015001}. One can guess that if a special Weyl pair composed of two different types of WPs is present, the chiral anomaly will switch the number of distinct types of Weyl fermions. In comparison with a conventional Weyl pair, it is expected that this unconventional Weyl pair may give rise to exotic transport phenomena since the intrinsic geometry around distinct types of WPs is rather different. However, unfortunately, an unconventional Weyl pair constructed by distinct types of WPs has not been reported in literatures.

As is well-known, due to twofold-degenerate features, WPs always appear in a material with either the parity ($\mathcal{P}$) or time-reversal ($\mathcal{T}$) symmetry broken. Therefore, besides the intrinsic WSMs, the Weyl fermions can also be obtained from other topological phases [e.g., Dirac semimetals (DSMs) and nodal-line semimetals (NLSMs)] by artificially breaking the related symmetries, such as strain, dopping, molecular adsorption, external magnetic fields, and light irradiation, etc. Among these approaches, the application of light irradiation is highly effective \cite{PhysRevLett.120.237403, PhysRevLett.120.156406, Wang453,PhysRevLett.117.090402,Sie1066,SentefNC, PhysRevLett.117.087402, PhysRevB.96.041206, PhysRevB.94.155206, PhysRevB.97.155152}. On the one hand, the breaking of specific symmetries can be conveniently controlled by the propagation or polarization direction of a incident light. On the other hand, the light irradiation produces fascinating Floquet-Bloch states \cite{Sie1066}, whose tunable band profiles depending on wavevector $\mathbf{k}$, facilitating to explore or design novel topological states of matter. The light irradiation not only paves the possibility to realize the unconventional Weyl pair composed of distinct types of WPs but also provides a reliable pathway for exploring desirable topological features with wide applications.

\begin{figure}
\setlength{\belowcaptionskip}{-0.2cm}
	\centering
	\includegraphics[scale=0.39]{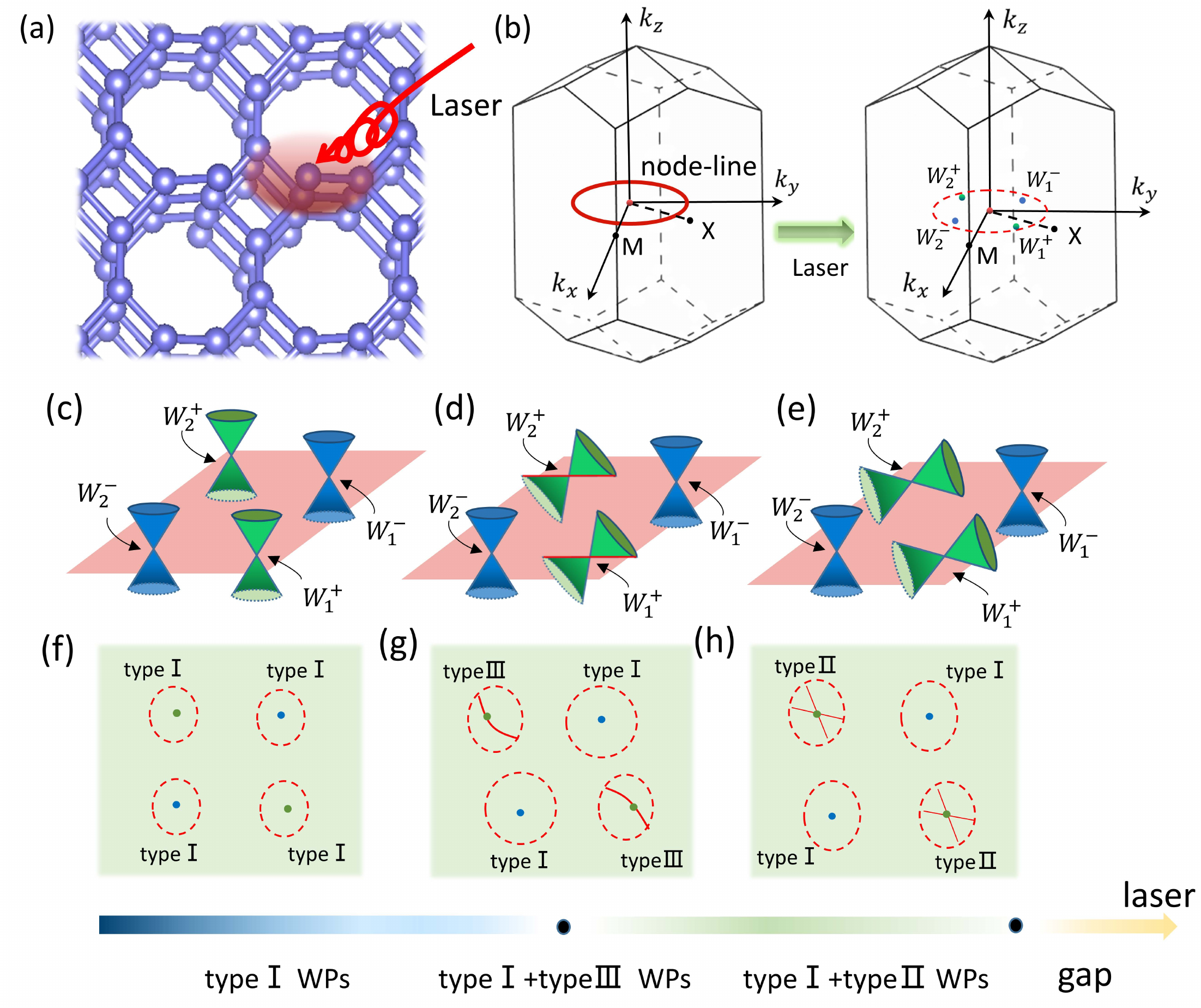}
	\caption{Light-modulated topological states in bct-C$_{16}$ driven by linearly polarized layer (LPL) $\mathbf{A}(\tau)=(0, 0, A_z\sin(\omega \tau))$. (a) Schematic figure of  bct-C$_{16}$ irradiated by incident laser $\mathbf{A}(\tau)$. (b) An ideal Dirac nodal-line located at the $k_x$-$k_y$ plane with $k_z = 0$ is transitioned to two pairs of WPs once the laser is applied. (c)-(e) The types of WPs evolve with increasing the light intensity. (f)-(h) The evolution of Fermi surface corresponds to (c)-(e), respectively. The band gap is present when the light intensity exceeds a critical value.
\label{figure1}}
\end{figure}

Here, based on first-principles calculations and the Floquet theorem, we demonstrate that the light-induced topological phase transition in a three-dimensional (3D) carbon phase can satisfy the criteria mentioned above. By using low-energy effective models, Weyl fermions have been investigated in several light-driven NLSMs \cite{PhysRevLett.117.087402, PhysRevB.96.041206, PhysRevB.94.155206, PhysRevB.97.155152}. However, the fascinating phenomena have rarely been realized in realistic materials. Recently, 3D carbon allotropes with topological-protected fermionic quasiparticles have been intensely investigated \cite{TSM-1,NLSM-bcoC16,Carbonnc, PhysRevLett.120.026402, PhysRevB.101.235119,TSM-bctC16}. Since the spin-orbital coupling (SOC) effect of a carbon element is negligible, the interplay between the SOC and light irradiation can be ignored. Therefore, a carbon allotrope can be considered as an ideal platform to study the photon-dressed topological states. In this work, we focus on the light-modulated topological states in bct-C$_{16}$, a typical NLSM protected by the parity-time reversal ($\mathcal{PT}$) symmetry \cite{TSM-bctC16,Ding_2020}. In ambient conditions, the carbon allotrope bct-C$_{16}$ crystallizes in a body-centered tetragonal (bct) structure with space group $I4_1/amd$ [see Fig. \ref{figure1}(a)], which can be obtained from the famous T-carbon through a temperature-driven structural transition \cite{Ding_2020}. As shown Fig. \ref{figure1}(b), our calculations indicate that bct-C$_{16}$ hosts an ideal Dirac nodal-line located at the the mirror reflection invariant $k_x$-$k_y$ plane with $k_z = 0$, which agrees well with the previous results \cite{TSM-bctC16,Ding_2020}. Under a periodic field of a linearly polarized laser (LPL), we show that the NLSM phase of bct-C$_{16}$ is transitioned to a mixed-WSM phase with two pairs of tunable WPs [see Fig. \ref{figure1}(b)].  With increasing the light intensity, right-handed WPs $W_1^{-}$ and $W_2^{-}$ are always preserved to type-I, while left-handed WPs $W_1^{+}$ and $W_2^{+}$ undergo a evolution from type-I to type-II [see Figs. \ref{figure1}(c)-\ref{figure1}(h)]. As a result, the coexistence of type-I and type-II WPs gives rise to the unconventional Weyl pairs composed of distinct types of WPs. During the transition process, the left-handed WPs can go through a critical type-III state, i.e., the Weyl pairs formed by type-I and type-III WPs can be present [see Figs. \ref{figure1}(d) and \ref{figure1}(g)].

To reveal the light-induced topological phase transition in bct-C$_{16}$, we carried out first-principles calculations to obtain the basis of plane waves as implemented in the Vienna ab initio simulation package \cite{Kresse2} [see the details in the Supplemental Material (SM) \cite{SM}]. By projecting plane waves of Bloch states onto the localized Wannier basis of C atoms using WANNIER90 package \cite{Mostofi2008,Marzari2012}, we constructed the Wannier tight-binding (TB) Hamiltonian as
\begin{equation}
H^{W}=\sum_{m,n,\mathbf{R},\mathbf{R}'}t_{mn}(\mathbf{R}-\mathbf{R}')C_{m}^{\dag}(\mathbf{R})C_{n}(\mathbf{R}')+h.c.,
\end{equation}
where $\mathbf{R}$ and $\mathbf{R}'$ are lattice vectors, ($m$,$n$) is the index of Wannier orbitals, $t_{mn}(\mathbf{R}-\mathbf{R}')$ are the hopping integrals between Wannier orbital $m$ at site $\mathbf{R}$ and Wannier orbital $n$ at site $\mathbf{R}'$, and $C_m^{\dag}(\mathbf{R})$ or $C_m (\mathbf{R})$ creates or annihilates an electron of Wannier orbital $m$ on site $\mathbf{R}$. When a time-periodic and space-homogeneous monochromatic laser field is applied to  bct-C$_{16}$ [see Fig. \ref{figure1}(a)], the time-dependent hopping integrals are obtained by using the Peierls substitution \cite{PhysRevA.27.72,PhysRevLett.110.200403}
\begin{equation}
t_{mn}(\mathbf{R}-\mathbf{R}', \tau)=t_{mn}(\mathbf{R}-\mathbf{R}')e^{i\frac{e}{\hbar}\mathbf{A}(\tau)\cdot \mathbf{d}_{mn}},
\end{equation}
where $\mathbf{A}(\tau)$ is the time-dependent vector potential of an applied laser-field, and $\mathbf{d}_{mn}$ is the related position vector between Wannier orbital $m$ at site $\mathbf{R}$ and Wannier orbital $n$ at site $\mathbf{R}'$. The corresponding light-driven operator is $C_{m}(\mathbf{R}, \tau)= \sum_{\alpha=-\infty}^{\infty} C_{\alpha m}(\mathbf{R})e^{i\alpha \omega \tau}$ with the Floquet operator $C_{\alpha m}(\mathbf{R})$ \cite{PhysRevLett.110.200403}.  In this case, the time-dependent $H^{W}(\tau)$ hosts both lattice and time translational symmetries, so we can map it onto a time-independent Hamiltonian according to the Floquet theory \cite{SM,PhysRevA.27.72,PhysRevLett.110.200403}. By carrying out a dual Fourier transformation, the static Floquet Hamiltonian can be expressed as
\begin{widetext}
\begin{equation}\label{eq3}
H^F({\mathbf{k}}, \omega)=\sum_{m, n}\sum_{\alpha, \beta}[H_{mn}^{\alpha-\beta}({\mathbf{k}}, \omega)+(\alpha-\beta)\hbar \omega \delta_{mn}\delta_{\alpha \beta}]C_{\alpha m}^{\dag}(\mathbf{k})C_{\beta n}(\mathbf{k})+h.c.,
\end{equation}
\end{widetext}
where $\omega$ is the frequency of an incident layer and thus $\hbar \omega$ represents the energy of photo, and the matrix $H_{mn}^{\alpha-\beta}({\mathbf{k}}, \omega)$ can be obtained by Wannier Hamiltonian as
\begin{widetext}
\begin{equation}\label{eq4}
H_{mn}^{\alpha-\beta}({\mathbf{k}}, \omega)=\sum_{\mathbf{R}}\sum_{\mathbf{R}'}e^{i\mathbf{k}\cdot (\mathbf{R}-\mathbf{R}')} \bigg(\frac{1}{T}\int_{0}^{T}t_{mn}(\mathbf{R}-\mathbf{R}')e^{i\frac{e}{\hbar}\mathbf{A}(\tau)\cdot \mathbf{d}_{mn}}e^{i(\alpha-\beta)\omega \tau}d\tau \bigg).
\end{equation}
\end{widetext}
Generally, the incident layer spans the Hilbert space of $H^F({\mathbf{k}}, \omega)$ to infinite dimensions, but the matrix $H_{mn}^{\alpha-\beta}({\mathbf{k}}, \omega)$ is damped rapidly with its order $|\alpha-\beta|$ increasing. Here, we truncate $H^F({\mathbf{k}}, \omega)$ to the second order ($|\alpha-\beta|=$ 0, 1, 2), which can accurately describe the photo-dressed band structures of bct-C$_{16}$ (see the SM \cite{SM}). In the main text, we focus on the results of a LPL. The results of a circularly polarized light (CPL) shown in the SM \cite{SM} indicate that the CPL also leads to a topological phase transition but does not result in a mixed-WSM phase.

\begin{figure}
\setlength{\belowcaptionskip}{-0.2cm}
\setlength{\abovecaptionskip}{-0.2cm}
	\centering
	\includegraphics[scale=0.49]{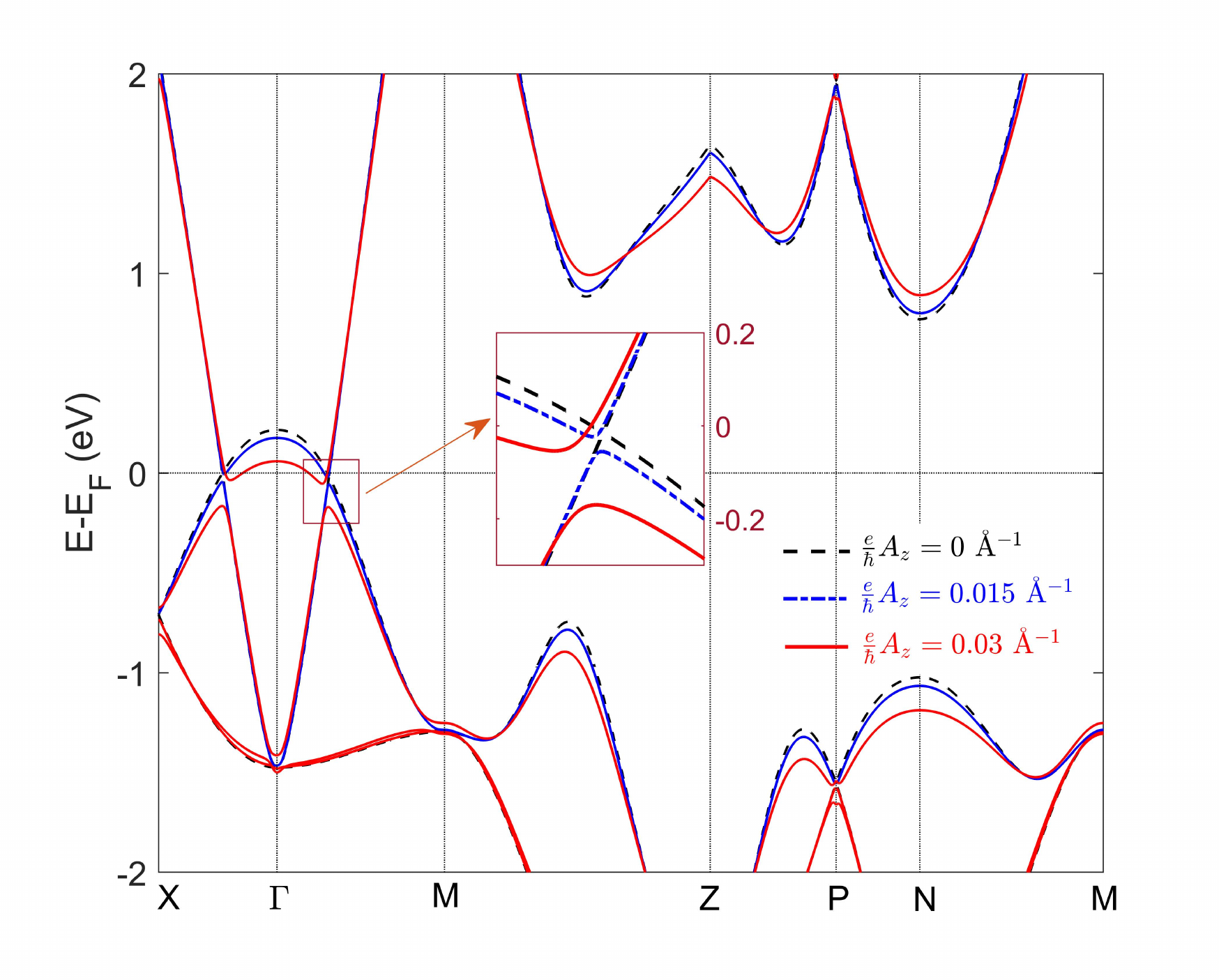}
	\caption{Floquet band structure evolution of bct-C$_{16}$ along the high-symmetry paths under the irradiation of a LPL. The black dashed lines, blue dot-dashed lines, and red solid lines represent a light intensity $eA_z /{\hbar}=$0.0, 0.015, 0.03 {\AA}$^{-1}$, respectively. The inset shows enlarged views around the original nodal point along $\Gamma$-$M$.
\label{figure2}}
\end{figure}

\begin{figure}
\setlength{\belowcaptionskip}{-0.3cm}
	\centering
	\includegraphics[scale=0.49]{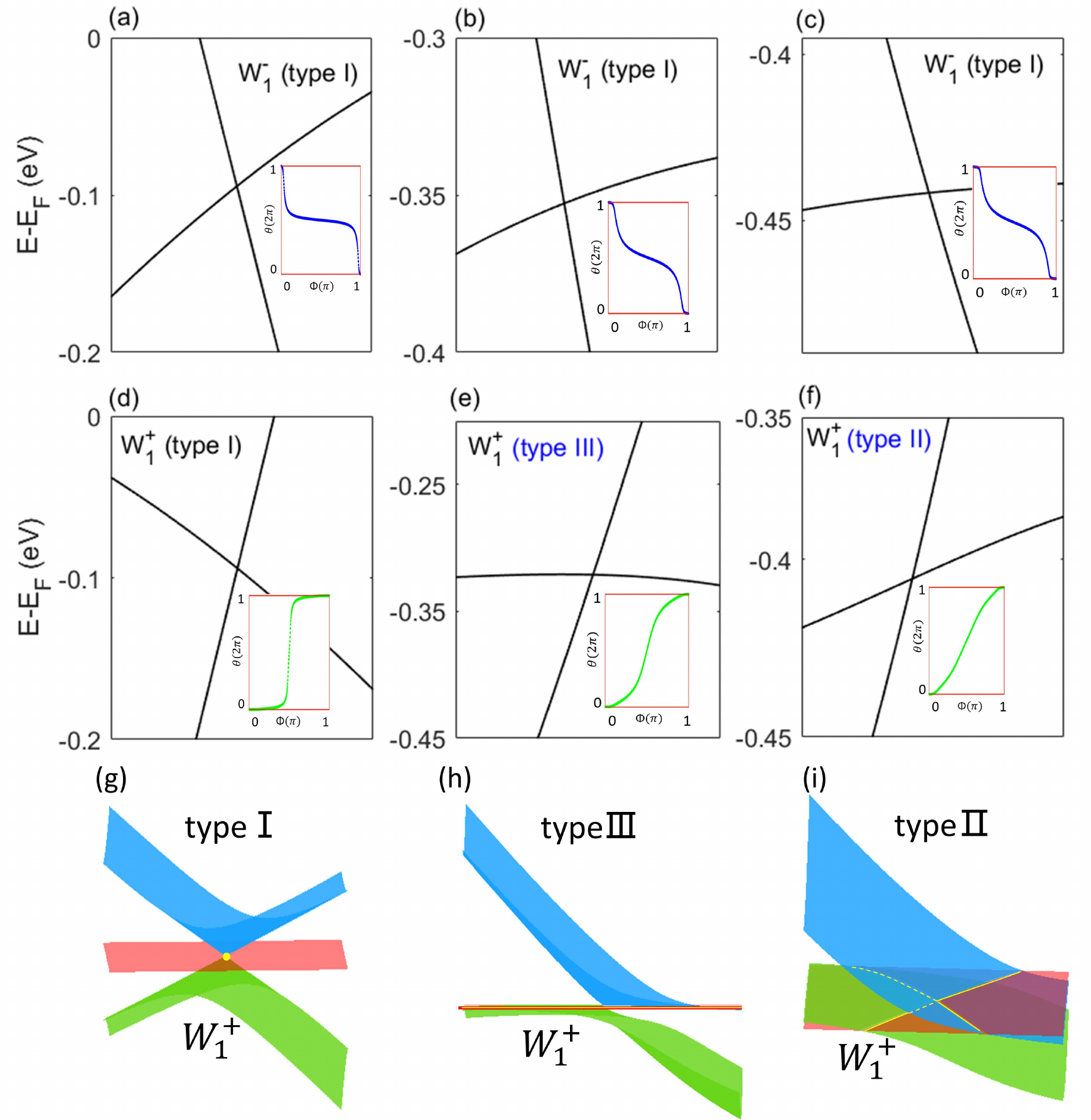}
	\caption{Band profiles around one pair of WPs $W_1^+$ and $W_1^-$ under the irradiation of a LPL with a light intensity (a), (d), and (g) $eA_z /{\hbar}=$0.03 {\AA}$^{-1}$, (b), (e), and (h) $eA_z /{\hbar}=$0.059 {\AA}$^{-1}$, and (c), (f) and (i) $eA_z /{\hbar}=$0.066 {\AA}$^{-1}$. Panels (a), (b), and (c) indicate that the right-handed WP $W_1^-$ is always kept to be type-I. Panels (c), (d), and (f) indicate that the left-handed WP $W_1^+$ undergos a transition from type-I to type-II, and a critical type-III WP is present at the boundary. The insets show the evolution of the Wannier charge centers around the WPs $W_1^-$ and $W_1^+$, respectively. Panels (g), (h), and (i) show the 3D plots of band dispersion around $W_1^+$, exhibiting standard type-I Weyl cone, type-III Weyl cone combining a flat band, and tilted type-II Weyl cone, respectively . The conduction and valence bands are respectively marked as blue and green. The isoenergetic contours corresponding to the Fermi surfaces are colored by yellow.
\label{figure3}}
\end{figure}

Next, we illustrate the topological phase transition of bct-C$_{16}$ under the light irradiation by diagonalization of the Floquet TB Hamiltonian Eq. (\ref{eq3}). To reveal the transition process, we employ a LPL with a time-periodic vector potential $\mathbf{A}(\tau)=(0, 0, A_z\sin(\omega \tau))$, where $A_z$ is its amplitude. The incident direction parallels to the $x$-$y$ plane, and the polarization is along the $z$-axis. In order to avoid the Floquet subbands crossing each other, we set the photo energy to $\hbar \omega =$ 5 eV, which is larger than the band width of bct-C$_{16}$. Once the light irradiation $\mathbf{A}(\tau)$ is applied, the wavevector will be coupled with vector potential by $k_z \rightarrow k_z+eA_z \sin(\omega \tau)/{\hbar}$. In this case, the $\mathcal{P}$-symmetry is broken, which may destroy the $\mathcal{TP}$-symmetry protected nodal-line in bct-C$_{16}$. As shown in Fig. \ref{figure2}, we compare band structures without light irradiation (black dashed lines) with those of light intensities  $eA_z /{\hbar}=$0.015 {\AA}$^{-1}$ (blue dot-dashed lines) and 0.03 {\AA}$^{-1}$ (red solid lines). One can see that the light irradiation obviously influences the electronic band structures of bct-C$_{16}$. As expected, the previous band crossings in the $\Gamma$-$X$ and $\Gamma$-$M$ directions are both gapped, indicating that the nodal-line fermions in bct-C$_{16}$ is disappeared. With increasing the light intensity, the band gaps are further enlarged (see the inset of Fig. \ref{figure2}). However, Fig. \ref{figure2} exhibits that the band inversion at the $\Gamma$ point is preserved. Hence, bct-C$_{16}$ under the light irradiation of a LPL $A_z\sin(\omega \tau)$ still maintains the nontrivial band topology. Through carefully checking energy differences between valence and conduction bands, we find that there are four nodal points in the whole BZ [see Fig. \ref{figure1}(b)]. The nodal points are below the Fermi level, making electron doping in bct-C$_{16}$. Each nodal point with specific chirality is slightly deviated from the crossing point of the original nodal line, forming two pairs of WPs: ($W_1^+$, $W_1^-$) and ($W_2^+$, $W_2^-$). In this case, the light-induced WSM phase of bct-C$_{16}$ exhibits excellent topological features with a minimum number of WPs in a $\mathcal{T}$-preserved system, in which the WPs with same chirality are symmetrically distributed with respect to the $\Gamma$ point, i.e., $\mathbf{k}_{W_1^+}=-\mathbf{k}_{W_2^+}$ and $\mathbf{k}_{W_1^-}=-\mathbf{k}_{W_2^-}$. Since the light-coupling in the BZ is momentum-dependent, the positions of WPs will evolve with the light amplitude $A_z$ . The coordinates of WPs in the momentum space at several typical light intensities are listed in the SM \cite{SM}. In addition, it is worth noting that there is a critical value $eA_z /{\hbar}=$0.069 {\AA}$^{-1}$. When a light intensity exceeds this critical value, all WPs annihilate with each other and bct-C$_{16}$ becomes a trivial insulator [see Fig. \ref{figure1}].

As depicted in Fig. \ref{figure2}, the light-induced modulation of band structures is dependent on the wavevector $\mathbf{k}$. Hence, low-energy excitations around different WPs may show different evolution behaviors under the light irradiation of a LPL. To better understand the photo-dressed Weyl fermions in bct-C$_{16}$, we present the band profiles around one pair of WPs (i.e., $W_1^+$ and $W_1^-$) evolving with increasing the amplitude of a LPL as shown in Fig. \ref{figure3}. The other pair of WPs (i.e., $W_2^+$ and $W_2^-$) shows the same behaviors with respect to the $\mathcal{T}$-symmetry. The band dispersion around the $W_1^-$ with a light intensity $eA_z /{\hbar}$ of 0.03, 0.059, and 0.066 {\AA}$^{-1}$ is illustrated in Figs. \ref{figure3}(a), \ref{figure3}(b), and \ref{figure3}(c), respectively. It is found that the right-handed WP $W_1^-$ is always kept to be type-I though it becomes more tilted with increasing the light intensity. On the contrary, light-dependent change around $W_1^+$ is more remarkable. When the light intensity $eA_z /{\hbar}$ increases from 0.03 to 0.066 {\AA}$^{-1}$, the left-handed WP $W_1^+$ undergos a transition from type-I [see Fig. \ref{figure3}(d)] to type-II [see Fig. \ref{figure3}(f)]. In this transition process, the critical type-III WP is present at the boundary between type-I and type-II states with $eA_z /{\hbar} = 0.059$ {\AA}$^{-1}$ [see Fig. \ref{figure3}(e)]. The 3D plot of band profiles around $W_1^+$ with $eA_z /{\hbar}$ = 0.03, 0.059, and 0.066 {\AA}$^{-1}$ is respectively shown in Figs. \ref{figure3}(g), \ref{figure3}(h), and \ref{figure3}(i), which are consistent with topologically nontrivial features of type-I, type-II, and type-III Weyl fermions. Besides, to ensure that the nodal-points under different light intensities are indeed WPs, we calculate the evolution of Wannier charge centers by employing the Wilson loop method \cite{Yu2011} [see the insets of Figs. \ref{figure3}(a)-\ref{figure3}(f)]. Our calculated results demonstrate that the light-induced mixed-WSM phase with two unconventional Weyl pairs composed of distinct types of WPs is realized in bct-C$_{16}$.

\begin{figure}
\setlength{\belowcaptionskip}{-0.2cm}
	\centering
	\includegraphics[scale=0.145]{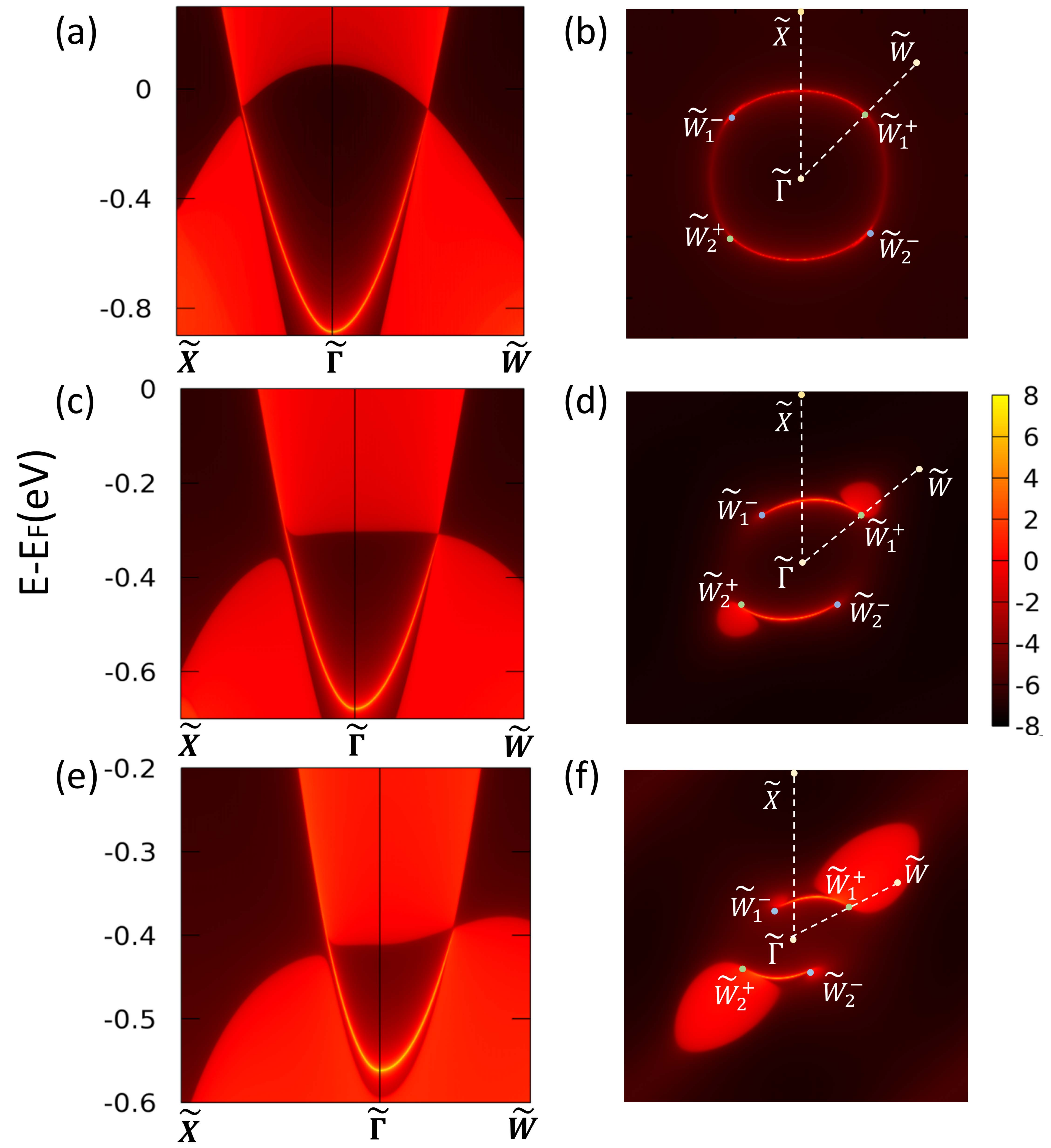}
	\caption{The calculated photo-dressed LDOS and Fermi surfaces projected on the semi-infinite (001) surface of bct-C$_{16}$ with a light intensity $eA_z /{\hbar}$ of (a), (b) 0.03, (c), (d) 0.059, and (e), (f) 0.066 {\AA}$^{-1}$. The green and blue dots denote the projected WPs with left-handed and right-handed chirality, respectively. In panels (b), (d), and (f), Floquet Fermi arcs connecting two projected WPs with opposite chirality are clearly visible. The pathes for LDOS in panels (a), (c), and (e) are marked in white-dashed lines in panels (b), (d), and (f).
\label{figure4}}
\end{figure}

One significant consequence of unconventional Weyl pairs in bct-C$_{16}$ is the existence of topologically protected surface states. To reveal this nontrivial properties, we calculate surface states using the iterative Green¡¯s method \cite{Sancho1984,WU2017} based on the Floquet TB Hamiltonian Eq. (\ref{eq3}). The calculated photo-dressed Fermi surfaces and local density of states (LDOS) projected on the semi-infinite (001) surface of bct-C$_{16}$ with a light intensity $eA_z /{\hbar}$ of 0.03, 0.059, and 0.066 {\AA}$^{-1}$ are respectively shown in Figs. \ref{figure4}(a)-\ref{figure4}(f). The LDOS show that there is a visible gap along $\tilde{\Gamma}$-$\tilde{X}$ and a projected Weyl cone with linear dispersion along $\tilde{\Gamma}$-$\tilde{W}$. Under different intensities, the projected band profiles around $\tilde{W}_1^+$ respectively exhibit type-I [Fig. \ref{figure4}(a)], type-III [ Fig. \ref{figure4}(c)], and type-II  [Fig. \ref{figure4}(e)] Weyl features. We can see that there are always two Floquet Fermi arcs connecting two projected WPs with opposite chirality (i.e, $\tilde{W}_1^+$ and $\tilde{W}_1^-$ or $\tilde{W}_2^+$ and $\tilde{W}_2^-$). Especially, the exotic Fermi arcs connecting two distinct types of WPs are present in Figs. \ref{figure4}(d) and \ref{figure4}(f). Besides, it is worth noting that separation between paired WPs decreases with increasing light intensities.

In conclusion, based on  first-principles calculations and the Floquet theorem, we propose that the Floquet mixed-WSM features with tunable WPs are present in a carbon allotrope bct-C$_{16}$ under a periodic field of a LPL. This exotic WSM phase is derived from a $\mathcal{PT}$-protected nodal-line since the light irradiation of LPL breaks the $\mathcal{P}$-symmetry, resulting in a minimum number of WPs in a $\mathcal{T}$-preserved system. With increasing the light intensity, left-handed WPs evolve from type-I into type-II while right-handed ones are always preserved to be type-I, realizing photo-dressed unconventional Weyl pairs composed of distinct types of WPs. During the transition process, a critical state possessing special Weyl pairs with type-I and type-III WPs also appear in bct-C$_{16}$. These unconventional Weyl pairs can be expected to exhibit an unknown effect related to the chiral anomaly in the presence of external electric and magnetic fields. A very interesting issue is that the Fermi arcs connects the projections of two different types of WPs, giving rise to exotic one-way dissipationless electronic propagation channel. Considering the extremely tiny SOC effect in carbon materials, bct-C$_{16}$ offers an ideal candidate to investigate light-induced mixed-WSMs with wide applications. Our work not only realizes the exotic unconventional Weyl pairs constructed by different types of WPs but also demonstrates that the light irradiation is a fascinating avenue for exploring desirable topological features.

~~~\\
~~~\\

This work was supported by the National Natural Science Foundation of China (NSFC, Grants No. 11974062, No. 11704177, and No. 11947406),
the Chongqing Natural Science Foundation (Grants No. cstc2019jcyj-msxmX0563), the Fundamental Research Funds for the Central Universities of China (Grants No. 2019CDXYWL0029, and No. 2020CDJQY-A057). \\
~~~\\


%


\begin{widetext}

\setcounter{figure}{0}
\setcounter{equation}{0}

\makeatletter

\makeatother
\renewcommand{\thefigure}{S\arabic{figure}}
\renewcommand{\thetable}{S\Roman{table}}
\renewcommand{\theequation}{S\arabic{equation}}

\begin{center}
	\textbf{
		\large{Supplemental Material for}}
	\vspace{0.2cm}
	
	\textbf{
		\large{
			``A Photoinduced Floquet Mixed-Weyl Semimetallic Phase in a Carbon Allotrope" }
	}
\end{center}

\section{Computational methods}
We carried out first-principles calculations as implemented in Vienna $ab$ $initio$ simulation package \cite{Kresse2s} within the framework of density-functional theory \cite{Kohn}. The projector augmented-wave method was used to treat the core-valence electron interactions \cite{Blochl}. The cutoff energy of plane waves was set to 400 eV, and the first Brillouin zone (BZ) was sampled by $12\times 12 \times 12$ Monkhorst-Pack grid \cite{Monkhorst}. The exchange-correlation functionals were described by the generalized gradient approximation within the Perdew-Burke-Ernzerhof formalism \cite{Perdew1}.  The forces of all atoms were relaxed until the force on each atom is less than 0.001 eV/A. The convergence condition of electron self-consistent circuit is $10^{-6}$ eV. The tight-binding (TB) Hamiltonian was constructed by using maximally localized Wannier functions (MLWF) methods by using the WANNIER90 package \cite{Mostofi2008s,Marzari2012s}. For each C atoms, we chosen ${p_x}$, ${p_y}$, and ${p_z}$ orbitals, and thus there were totally 24 orbitals for one primitive unit cell. By projecting plane waves of Bloch states onto the localized Wannier basis of C atoms, we constructed the Wannier tight-binding (TB) Hamiltonian as
\begin{equation}
H^{W}=\sum_{m,n,\mathbf{R},\mathbf{R}'}t_{mn}(\mathbf{R}-\mathbf{R}')C_{m}^{\dag}(\mathbf{R})C_{n}(\mathbf{R}')+h.c.,
\end{equation}
where $\mathbf{R}$ and $\mathbf{R}'$ are lattice vectors, ($m$,$n$) is the index of Wannier orbitals, $t_{mn}(\mathbf{R}-\mathbf{R}')$ are the hopping integrals between Wannier orbital $m$ at site $\mathbf{R}$ and Wannier orbital $n$ at site $\mathbf{R}'$, and $C_m^{\dag}(\mathbf{R})$ or $C_m (\mathbf{R})$ creates or annihilates an electron of Wannier orbital $m$ on site $\mathbf{R}$.

\section{Implementation of the Floquet theory in Wannier TB model}
When a time-periodic and space-homogeneous monochromatic laser field $\mathbf{A}(\tau)$ is applied, the light-driven system is as a function of time $\tau$.  The time-dependent hopping integrals are obtained by using the minimal coupling in the inverse Fourier transform (i.e, Peierls substitution) \cite{PhysRevA.27.72s,PhysRevLett.110.200403s}
\begin{equation}
t_{mn}(\mathbf{R}-\mathbf{R}', \tau)=t_{mn}(\mathbf{R}-\mathbf{R}')e^{i\frac{e}{\hbar}\mathbf{A}(\tau)\cdot \mathbf{d}_{mn}},
\end{equation}
where $\mathbf{d}_{mn}$ is the related position vector between Wannier orbital $m$ at site $\mathbf{R}$ and Wannier orbital $n$ at site $\mathbf{R}'$. The light-driven operator can be expressed as
\begin{equation}
\begin{split}
&C_{m}(\mathbf{R}, \tau)= \sum_{\alpha=-\infty}^{\infty} C_{\alpha m}(\mathbf{R})e^{i\alpha \omega \tau}, \\
&C_{m}^{\dag}(\mathbf{R}, \tau)= \sum_{\alpha=-\infty}^{\infty} C_{\alpha m}^{\dag}(\mathbf{R})e^{-i\alpha \omega \tau},
\end{split}
\end{equation}
in which $C_{\alpha m}^{\dag}(\mathbf{R})$ or $C_{\alpha m}(\mathbf{R})$ is the creation or annihilation operator with the Floquet band index $\alpha$. In this case, the time-dependent Hamiltonian $H^{W}(\tau)$ hosts both lattice and time translational symmetries,i.e.,
\begin{equation}
H^{W}(\mathbf{r}+\mathbf{R}, \tau+T)=H^{W}(\mathbf{r}+\mathbf{R}, \tau)=H^{W}(\mathbf{r}, \tau+T),
\end{equation}
which is characterized by lattice vectors $\mathbf{R}$ and time period $T=\omega/2\pi$. Under these assumptions, we can use the Fourier transformations
to obtain the time-dependent Hamiltonian, which can be expressed as
\begin{equation}\label{Ht}
H(\mathbf{k},t)=\sum_{m,n}t_{mn}(\mathbf{R}-\mathbf{R}', \tau)C_{m}^{\dag}(\mathbf{k}, \tau)C_{n}(\mathbf{k}, \tau)+h.c.
\end{equation}
with
\begin{equation}
\begin{split}
&C_{m}(\mathbf{k}, \tau)=\sum_{\mathbf{R}}\sum_{\alpha=-\infty}^{\infty} C_{\alpha m}(\mathbf{R})e^{-i\mathbf{k}\cdot \mathbf{R} +  i\alpha \omega \tau},\\
&C_{m}^{\dag}(\mathbf{k}, \tau)=\sum_{\mathbf{R}}\sum_{\alpha=-\infty}^{\infty} C_{\alpha m}^{\dag}(\mathbf{R})e^{+i\mathbf{k}\cdot \mathbf{R} -  i\alpha \omega \tau}.
\end{split}
\end{equation}
Based on the Floquet theory, the static Floquet Hamiltonian can be obtained from the average of Eq. (\ref{Ht}) with respect to time $\tau$. As a result, the Hamiltonian as functions of wavevector $\mathbf{k}$ and frequency $\omega$ is expressed by
\begin{equation}
H^F({\mathbf{k}}, \omega)=\sum_{m, n}\sum_{\alpha, \beta}[H_{mn}^{\alpha-\beta}({\mathbf{k}}, \omega)+(\alpha-\beta)\hbar \omega \delta_{mn}\delta_{\alpha \beta}]C_{\alpha m}^{\dag}(\mathbf{k})C_{\beta n}(\mathbf{k})+h.c.,
\end{equation}
where $\hbar \omega$ represents the energy of photo, and the matrix $H_{mn}^{\alpha-\beta}({\mathbf{k}}, \omega)$ is
\begin{equation}\label{eqs6}
H_{mn}^{\alpha-\beta}({\mathbf{k}}, \omega)=\sum_{\mathbf{R}}\sum_{\mathbf{R}'}e^{i\mathbf{k}\cdot (\mathbf{R}-\mathbf{R}')} \bigg(\frac{1}{T}\int_{0}^{T}t_{mn}(\mathbf{R}-\mathbf{R}')e^{i\frac{e}{\hbar}\mathbf{A}(\tau)\cdot \mathbf{d}_{mn}}e^{i(\alpha-\beta)\omega \tau}d\tau \bigg).
\end{equation}
Generally, a incident light spans the Hilbert space of $H^F({\mathbf{k}}, \omega)$ to infinite dimensions shown in Fig. \ref{FIG-S1}.

\begin{figure}
\setlength{\belowcaptionskip}{-0.2cm}
	\centering
	\includegraphics[scale=0.5]{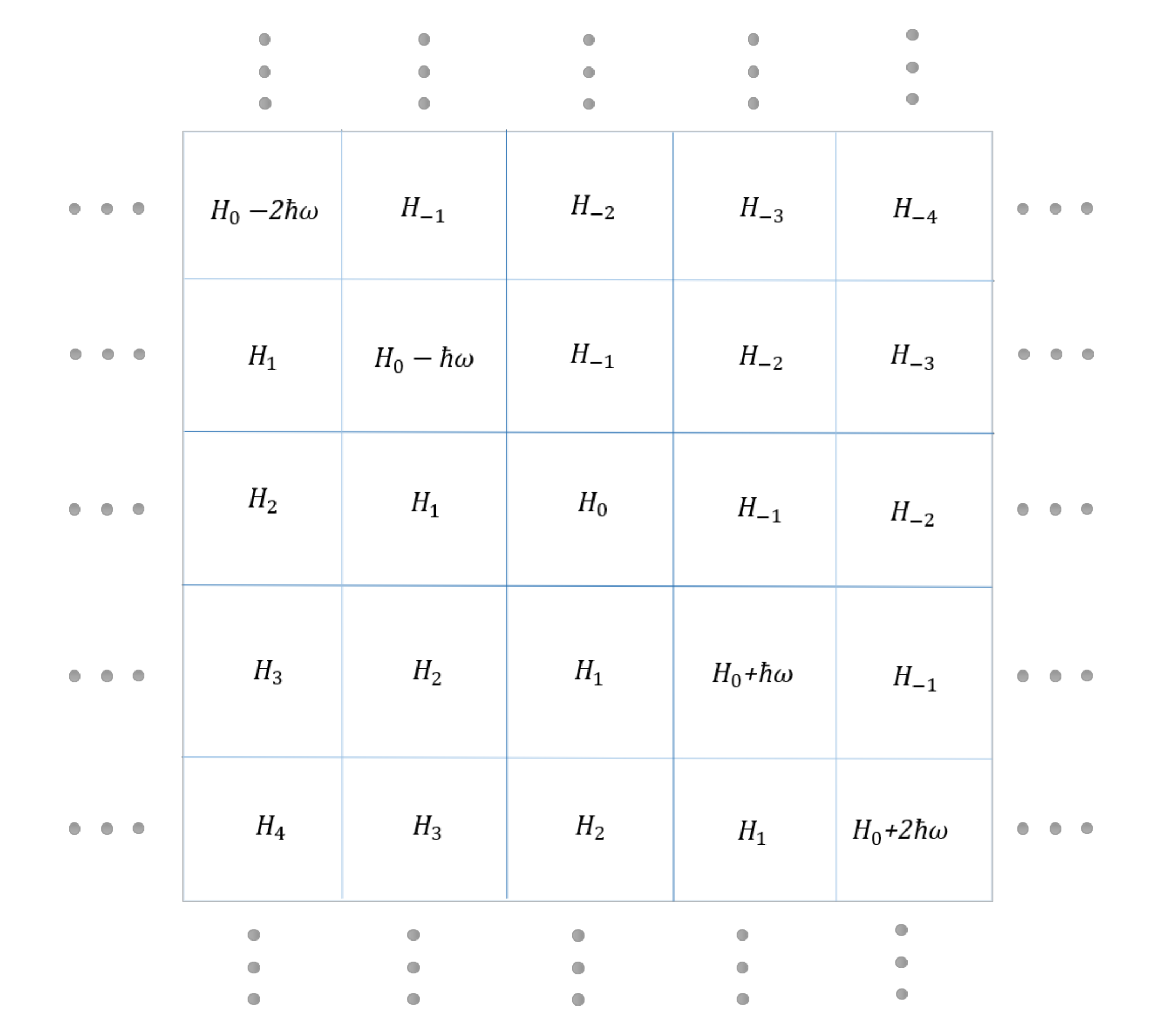}
	\caption{Infinite dimensions of Floquet Hamiltonian.
\label{FIG-S1}}
\end{figure}
To calculate the matrix $H_{mn}^{\alpha-\beta}({\mathbf{k}}, \omega)$ in Eq. (\ref{eqs6}), we give ${\bf{A}}(\tau )$ and ${{\bf{d}}_{{{mn}}}}$ as the following general forms
\begin{equation}
{\bf{A}}(\tau ) = [{A_x}\sin (\omega \tau  + {\varphi _1}),{A_y}\sin (\omega \tau  + {\varphi _2}),{A_z}\sin (\omega \tau  + {\varphi _3})],
\end{equation}
\begin{equation}
{{\bf{d}}_{mn}} = ({d_x},{d_y},{d_z}),
\end{equation}
where ${A_x}$, ${A_y}$, and ${A_z}$ represent the amplitude of potential vector along the $x$, $y$, and $z$ direction, respectively, and ${\varphi _1}$, ${\varphi _2}$, and ${\varphi _3}$ are the related initial phases.
Then, we can write the matrix element ${H_{mn}^q} (\mathbf{k}, \omega)$ ($q=\alpha-\beta$) as
\begin{equation}\label{eq4}
H_{mn}^q ({\mathbf{k}}, \omega)= \sum\limits_{\bf{R}} {\sum\limits_{{{\bf{R}}^{\bf{'}}}} {} } {t_{mn}}({\bf{R - }}{{\bf{R}}^{\bf{'}}}){e^{i{\bf{k}} \cdot ({\bf{R - }}{{\bf{R}}^{\bf{'}}})}}{e^{iq\varphi }} \cdot {J_q}\bigg(\frac{e}{\hbar }{A_{\max }}\bigg),
\end{equation}
in which $J_q$ is q-th Bessel function
\begin{equation}
{J_q}\bigg(\frac{e}{\hbar }{A_{\max }}\bigg)= {e^{-iq\varphi }} \frac{1}{T}  \int_0^T {{e^{i[\frac{e}{\hbar }{A_{\max }}\sin (\omega \tau  + \varphi )]}}{e^{iq\omega \tau }}d\tau }
\end{equation}
with
\begin{equation}
\begin{split}
&{A_{\max }} = \sqrt {{{\big({A_x}{d_x}\sin {\varphi _1} + {A_y}{d_y}\sin {\varphi _2}+{A_z}{d_z}\sin {\varphi _3}\big)}^2} + {{\big({A_x}{d_x}\cos {\varphi _1} + {A_y}{d_y}\cos {\varphi _2}+{A_z}{d_z}\cos {\varphi _3}\big)}^2}},\\
&\varphi  =  \arctan \bigg(\frac{{{A_x}{d_x}\sin {\varphi _1} + {A_y}{d_y}\sin {\varphi _2}+{A_z}{d_z}\sin {\varphi _3}}}{{{A_x}{d_x}\cos {\varphi _1} + {A_y}{d_y}\cos {\varphi _2}+{A_z}{d_z}\cos {\varphi _3}}}\bigg).
\end{split}
\end{equation}

Though the Hilbert space of $H^F({\mathbf{k}}, \omega)$ hosts infinite dimensions, but the matrix $H_{mn}^{q}({\mathbf{k}}, \omega)$ is damped rapidly to zero with its order $|q|$ increased. In this work, we truncate $H^F({\mathbf{k}}, \omega)$ to the second order ($q=$ 0, $\pm 1$, $\pm 2$), which can accurately describe the photo-dressed band structures of bct-C$_{16}$ as shown in Fig \ref{FIG.S2.}.

\begin{figure}
\setlength{\belowcaptionskip}{-0.2cm}
	\centering
	\includegraphics[scale=0.6]{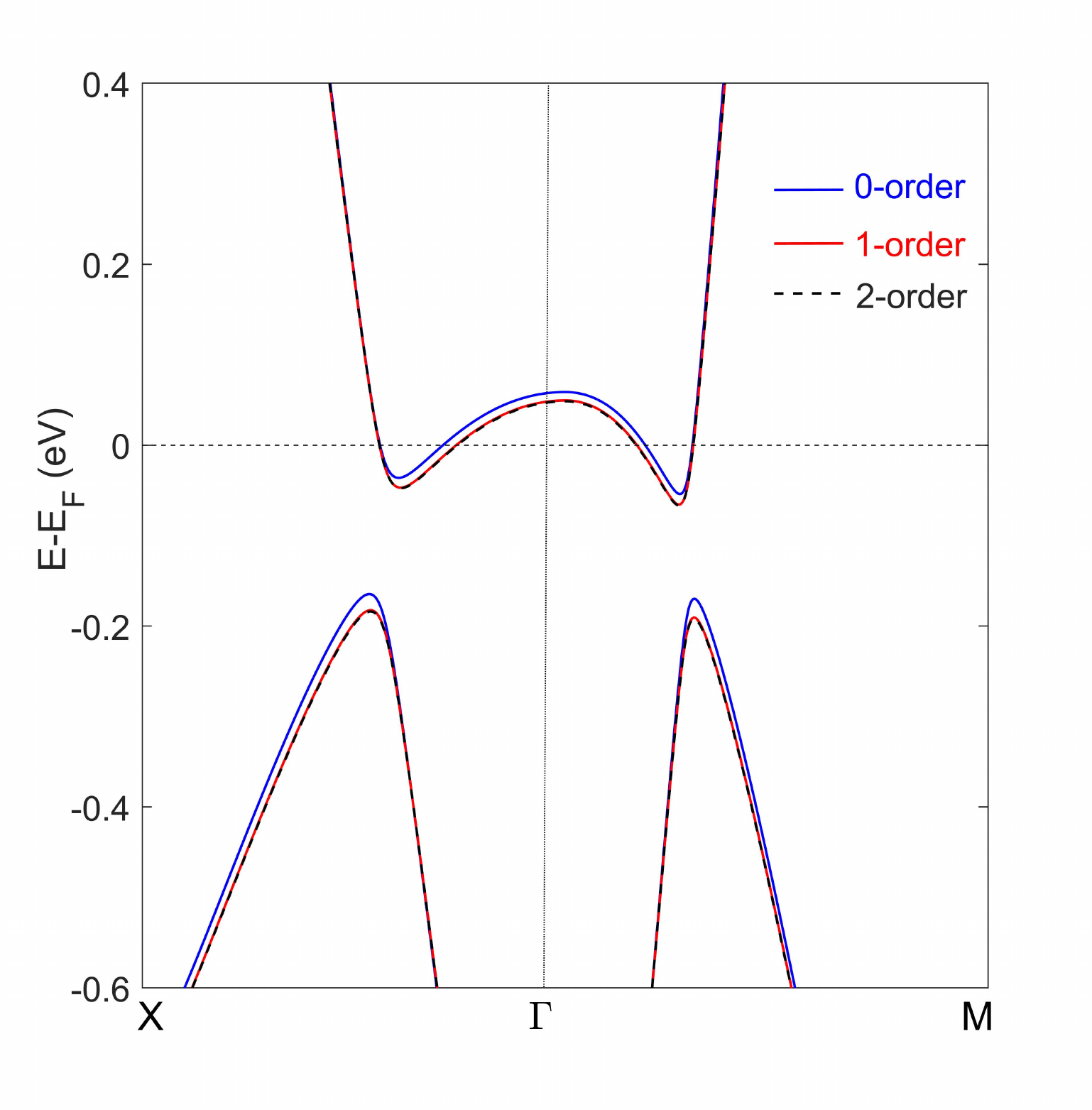}
	\caption{The comparison of Floquet band structures among different truncated orders q=0, 1, 2. We set $e{A_z}/\hbar $= 0.03 {\AA}$^{-1}$.
\label{FIG.S2.}}
\end{figure}

\section{The results of a circularly polarized light}
A circularly polarized light can also lead to a phase transition from a nodal-line to two pairs of Weyl points (WPs). However, we don't observe a mixed-WSM phase with unconventional Weyl pairs composed of distinct types of WPs. For instance, we list the WPs with a circularly polarized light $\mathbf{A}(\tau)=[A_x\sin(\omega \tau)),A_y\sin(\omega \tau+\pi/2), 0]$ in Table \ref{Tablec1}.
\begin{table}[!htbp]
\caption{The positions of WPs with a circularly polarized light. Here, we set $e{A_x}/\hbar $= $e{A_y}/\hbar $= 0.01{\AA}$^{-1}$. The coordinates of WPs in momentum space and the energies relative to $E_F$ are listed, respectively.}
	\begin{center}
	\begin{tabular}{p{1.2 cm}|p{5 cm}|*{1}{p{2 cm}} *{1}{p{1.2cm}} }
		\hline
\hline
		WP &\centering{[${k_x}$ ({\AA}$^{-1}$), ${k_y}$ ({\AA}$^{-1}$), ${k_z}$ ({\AA}$^{-1}$)]} &${E-E_F}$ (eV)\\
       \hline
		$W_1^ +$&\centering{(0.2722, -0.0967, -0.0046)}& -0.1600\\
		
		$W_1^ -$&\centering{(-0.2510, 0.0022, 0.0065)}& -0.1559\\
		
		$W_2^ +$&\centering{(-0.2722, 0.0967, 0.0046)}& -0.1600\\
		
		$W_2^ -$&\centering{(0.2510, -0.0022, -0.0065)} &-0.1559\\
		\hline
\hline
	\end{tabular}\label{Tablec1}
\end{center}
\end{table}

\newpage

\section{The positions of WPs under different light intensities}
Since the light-coupling is momentum-dependent, the positions of WPs will evolve with the light amplitude. Here, we list the coordinates of WPs in momentum space at several typical light intensities of a linearly polarized light in Table \ref{Tablel1}.
\begin{table}
\caption{The coordinates of WPs in momentum space at several typical light intensities of a linearly polarized light are listed. Here, we set $e{A_z}/\hbar $= 0.03, 0.059, 0.066 {\AA}$^{-1}$, respectively. The coordinates in momentum space and the energies relative to $E_F$ are listed, respectively.}
	\begin{tabular}{p{3 cm}|p{1.2 cm}|*{1}{p{5 cm}} *{1}{p{2 cm}} }
		\hline
\hline
	Intensity	& WP &\centering{[${k_x}$ ({\AA}$^{-1}$), ${k_y}$ ({\AA}$^{-1}$), ${k_z}$ ({\AA}$^{-1}$)]} &${E-E_F}$ (eV)\\
       \hline
 $e{A_z}/\hbar $= 0.03 {\AA}$^{-1}$   &$W_1^ +$&\centering{(0.2172, 0.1971, -0.0333)} &-0.0942\\
		
		&$W_1^ -$&\centering{(-0.2077, 0.1931, 0.0040)}&-0.0944\\
		
		&$W_2^ +$&\centering{(-0.2172, -0.1971, 0.0333)}&-0.0942\\
		
		&$W_2^ -$&\centering{(0.2077, -0.1931, -0.0040)}&-0.0944\\
		\hline
$e{A_z}/\hbar $= 0.059 {\AA}$^{-1}$   &$W_1^ +$&\centering{(0.1618,  0.1662, -0.1024)} &-0.3211\\
		
		&$W_1^ -$&\centering{(0.1084, -0.1366, 0.0045)}&-0.3527\\
		
		&$W_2^ +$&\centering{(-0.1618, -0.1662, 0.1024)}&-0.3211\\
		
		&$W_2^ -$&\centering{(-0.1084, 0.1366, -0.0045)}&-0.3527\\
\hline
$e{A_z}/\hbar $= 0.066 {\AA}$^{-1}$   &$W_1^ +$&\centering{(0.1203, 0.1351, -0.1045)} &-0.4054\\
		
		&$W_1^ -$&\centering{(0.0486, -0.0975, 0.0173)}&-0.4417\\
		
		&$W_2^ +$&\centering{(-0.1203, -0.1351, 0.1045)}&-0.4054\\
		
		&$W_2^ -$&\centering{(-0.0486, 0.0975, -0.0173)}&-0.4417\\
\hline
\hline
	\end{tabular}\label{Tablel1}
\end{table}

\section{The Fermi surfaces on two-dimensional cuts}
In Figs. \ref{FIG.S4}(a)-\ref{FIG.S4}(d), the Fermi surfaces of $e{A_z}/\hbar $= 0.05{\AA}$^{-1}$ on a two-dimensional cut are shown, confirming that two pairs of Weyl points are type-I. In Figs. \ref{FIG.S4}(e)-\ref{FIG.S4}(h), the Fermi surfaces of $e{A_z}/\hbar $= 0.0675{\AA}$^{-1}$ on a two-dimensional cut are shown, confirming that there are two type-I Weyl points (i.e., $W_1^ -$ and $W_2^ -$), and there are two type-II Weyl points (i.e., $W_1^ +$ and $W_2^ +$).
\begin{figure}
	\centering
	\includegraphics[scale=0.8]{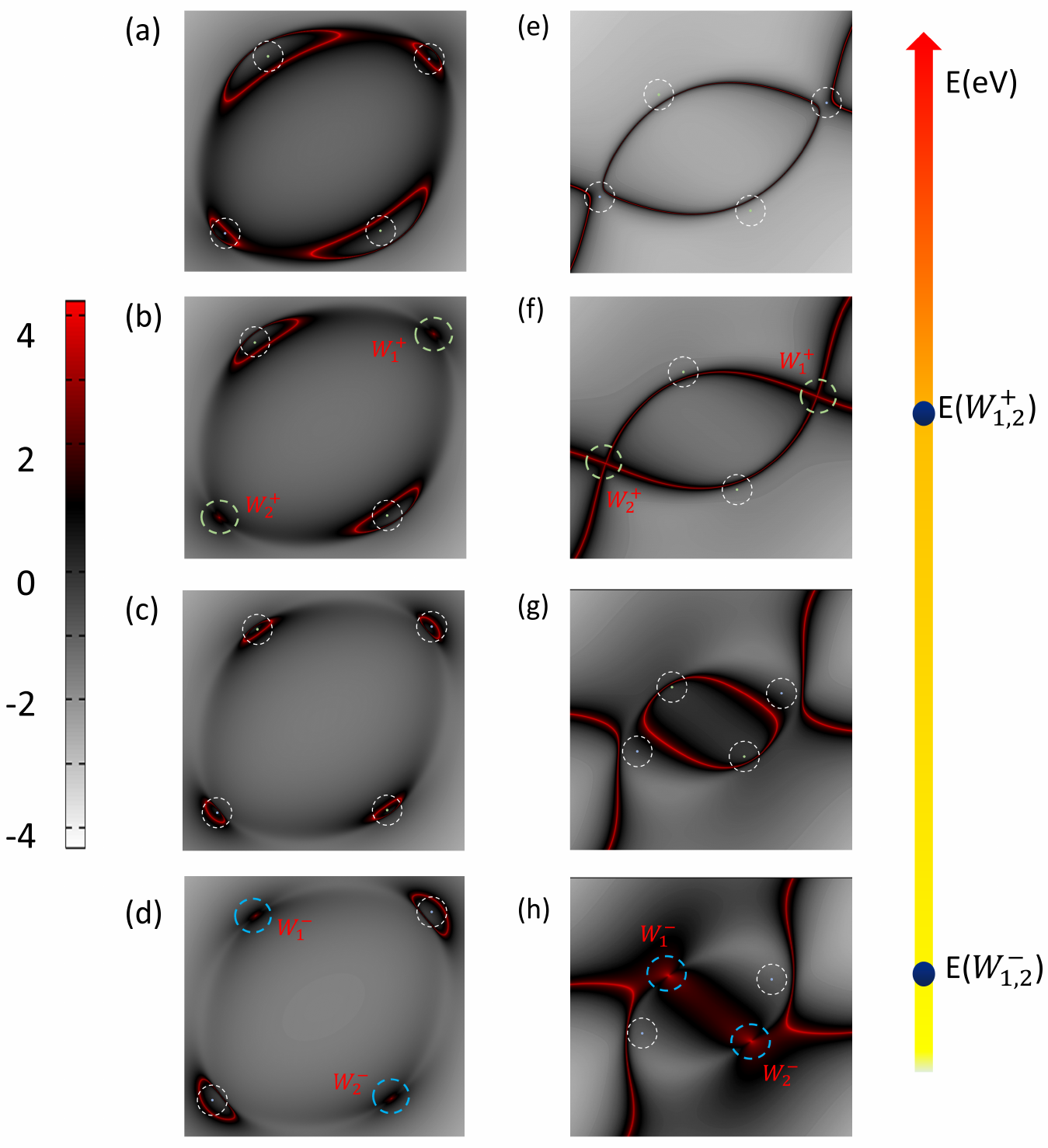}
	\caption{(a)-(d) The Fermi surfaces of $e{A_z}/\hbar $= 0.05{\AA}$^{-1}$. (e)-(h) The Fermi surfaces of $e{A_z}/\hbar $= 0.0675{\AA}$^{-1}$.
\label{FIG.S4}}
\end{figure}


%
\end{widetext}

\end{document}